\documentclass[twocolumn,showpacs,preprintnumbers,amsmath,amssymb]{revtex4}
\usepackage{graphicx}% Include figure files
\usepackage{dcolumn}% Align table columns on decimal point
\usepackage{bm}% bold math

\begin{document}

\preprint{APS/123-QED}

\title{Atomic diffraction in counter-propagating Gaussian pulses of laser light}
\author{Tapio P. Simula$^1$, Atom Muradyan$^2$ and Klaus M\o lmer$^1$}
\affiliation{$^1$Lundbeck Foundation Theoretical Center for Quantum System Research, \\
\mbox{Department of Physics and Astronomy, University of Aarhus, DK-8000 \AA rhus C, Denmark} \\
$^2$Department of Physics, Yerevan State University, 1 Alex Manukian, Yerevan 375049, Armenia}

\begin{abstract}
We present an analysis of atomic diffraction due to the interaction
of an atomic beam with a pair of Gaussian light pulses. We derive a
simple analytical expression for the populations in different
diffraction orders. The validity of the obtained solution extends
beyond the  Raman-Nath regime, where the kinetic energy associated
with different diffraction peaks is neglected, into the so-called
channeling regime where accurate analytical expressions have not
previously been available for the diffraction. Comparison with
experimental results and exact numerical solutions demonstrate the
validity of our analytical formula.
\end{abstract}


\maketitle

\section{Introduction}

Diffraction of light and matter waves lies at the heart of a number of diagnostic measurement
techniques in physics and engineering. Due to their internal level structure, atoms and molecules can
be diffracted by resonant or near resonant laser beams
\cite{earlyref1,earlyref2,earlyref3,earlyref4,earlyref5,earlyref6}. The field of atom optics points to
applications for high precision probing of inertial effects and fundamental physics
\cite{useofdiff1,useofdiff2,useofdiff3,useofdiff4}, as well as to the use of diffraction to diagnose
single- and multi-particle properties of the atoms themselves in various physical settings
\cite{manybody1,manybody2,manybody3}. The use of light induced potentials for structured deposition of
atoms on surfaces has also been demonstrated \cite{deposition}. By use of Laguerre-Gaussian laser
beams, atomic diffraction was recently used to transfer not only linear but also angular momentum to
an atomic gas \cite{Andersen2007a,Simula2007a}. By tailoring the polarization and frequencies of
light, it is possible to induce complicated multi-level dynamics with various potential applications
\cite{theoryreview1}. For most species, using optical pumping and suitable polarization schemes, it is
also possible to restrict the dynamics and to obtain a situation of effective propagation of a single
component wave function in a single periodic potential \cite{theoryreview1,theoryreview2}.

The coupling of atomic momentum components by absorption and stimulated emission events to a comb of
other momentum states separated by twice the photon momentum cannot be solved analytically due to the
non-commuting kinetic and potential energies. However, for two parameter regimes simple analytical
solutions have been known for a long time. In the Raman-Nath regime of short pulse durations, the
difference in kinetic energy of different momentum states is neglected and the coupling leads to an
expansion with Bessel-function expression for different momentum component amplitudes
\cite{theoryreview1,theoryreview2}. In the Bragg regime of long and weak pulses, kinetic energy must
be strictly conserved by absorption and emission of an integer number of counter-propagating photon
pairs, and the system reduces to an effective, analytically tractable, two-level system.

Atomic diffraction is applied extensively in experiments, and the need for numerical and analytical
theoretical approaches beyond the range of validity of the Raman-Nath and the Bragg theoretical
treatments, is explicitly expressed in the recent literature \cite{Keller1999a,Muller2007a}. Here we
depart from the Raman-Nath approach, and suggest an analytical formula which extends into the
channeling regime. Our analytical expression, includes effects due to the kinetic energy dispersion,
and deploying numerical simulations we show that it extends the range of validity of the simple
Raman-Nath approach into the channeling regime where both Raman-Nath and Bragg descriptions fail. In
Section II, we introduce our analytical and improved approximate theories. In Sec. III we present
numerical calculations and compare them to the different analytical approaches derived in the previous
section, assessing their validity regime. Finally, Sec. IV concludes the paper with a discussion.

\section{Theory}

\begin{figure}[t!] %  figure placement: here, top, bottom, or page
   \centering
   \includegraphics[width=8cm]{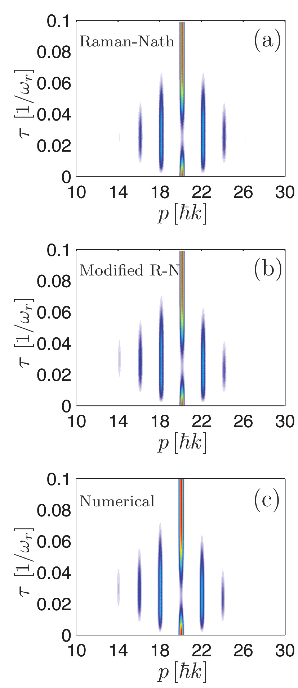}
   \caption{(Color online) Diffraction patterns for various pulse durations, $\tau$, as predicted by (a) the Raman-Nath solution
   Eq.~(\ref{eq5}), (b) its modified version, Eq.~(\ref{eq6}) and (c) the full numerical solution to the Eq.~(\ref{eq2}), plotted as functions of the momentum, $p$.
   The color indicates the population intensity such that red corresponds to the maximum and dark blue to the minimum. Parameters are: $q=52.9$ and $p_0=20\hbar k$. }
\label{F1}
\end{figure}

\begin{figure}[t!] %  figure placement: here, top, bottom, or page
   \centering
   \includegraphics[width=8.6cm]{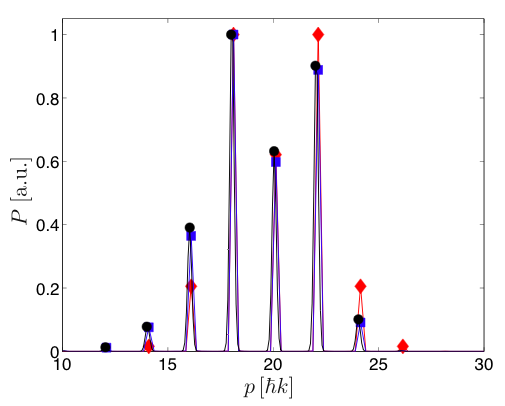}
       \caption{(Color online) Diffraction intensities for a pulse duration $ \tau=0.0375/\omega_r$ and intensity $q=52.9$ in reduced units.
       The diffraction probabilities are shown as function of the final state momentum.
       Red diamonds, blue boxes and black bullets correspond to the frames (a), (b), and (c) in Fig.~\ref{F1},
       respectively.}
\label{F2}
\end{figure}

We consider an atom with mass $M$, two atomic levels with energy level separation $\hbar \omega_0$, and
dipole matrix element $d$, interacting with two counter-propagating laser field pulses,
\begin{equation}
\mathcal{E}(z,t)=\mathcal{E}_1(t-z/c) e^{i(kz-\omega t)}+\mathcal{E}_2(t+z/c)e^{-i(kz+\omega t)} + c.c.
\end{equation}
We imagine that the two pulses arrive at the origin, $z=0$,
symmetrically from opposite directions. For realistic atomic and
field parameters, the width of the atomic distribution around $z=0$
will be narrow compared to the spatial range of intensity variation
of either light pulse and hence their interaction with the atoms is
that of a standing wave with a time dependent strength. We assume
that the atoms have initially a well defined momentum, $p_0$, and we
adiabatically eliminate the excited atomic state, which is
off-resonance from the laser fields by an amount
$\Delta=\omega-\omega_0$, yielding the Schr\"odinger equation for
the ground state wave function, $\psi(z,t)\equiv
a(z,t)\exp(ip_0z/\hbar-ip_0^2 t/2M\hbar)$,
\begin{equation}
i\hbar\frac{\partial}{\partial t}a(z,t)=\left( -\frac{\hbar^2}{2M}\frac{\partial^2}{\partial z^2} + i\hbar\frac{p_0}{M}\frac{\partial}{\partial z} + V_{\rm ext}(z,t)\right)a(z,t)
\label{eq2}
\end{equation}
where
\begin{eqnarray}
V_{\rm ext}(z,t)&=&\frac{|d|^2}{\hbar\Delta} \Big(|\mathcal{E}_1(t-z/c)|^2+|\mathcal{E}_2(t+z/c)|^2 \notag\\
&+&2\textrm{Re}(\mathcal{E}_1^*(t-z/c)\mathcal{E}_2(t+z/c)e^{-2ikz}\Big)
\end{eqnarray}

\subsection{Neglecting the variation of the kinetic energy for different momentum states}

Neglecting the second order spatial derivative in Eq.~(\ref{eq2}), it becomes solvable by
straightforward integration \cite{Muradyan}. We replace the assumed smooth position dependence of the
amplitude, $a(z,t)$, before the interaction with the laser field by a constant, arriving at the solution
\begin{eqnarray}
a(z,t) &\propto& \exp \Bigg(  \frac{i|d|^2}{\hbar^2\Delta}  \Big(  \int_{-\infty}^t |\mathcal{E}_1|^2\rm{d}t' +\int_{-\infty}^t |\mathcal{E}_2|^2\rm{d}t' \notag\\
&+&e^{i2k(p_0t/M-z)}\int_{-\infty}^t|\mathcal{E}_1|^2|\mathcal{E}_2|^2e^{-i2kp_0 t'/M}\rm{d}t'\notag \\
&+&e^{-i2k(p_0t/M-z)}\int_{-\infty}^t|\mathcal{E}_1|^2|\mathcal{E}_2|^2e^{i2kp_0 t'/M}\rm{d}t' \Big) \Bigg),
\label{eq4}
\end{eqnarray}
where we will assume real field amplitudes. Physically, neglecting
the kinetic energy operator amounts to making the dynamics local in
position involving only a position dependent phase factor. Assuming
Gaussian pulses, $\mathcal{E}_{1,2}(t\pm z/c)=E_{1,2}\exp\{-(t\pm
z/c)^2/\tau^2\}$, the definite integrals in Eq.~(\ref{eq4}) can be
evaluated at $t=\infty$, i.e., after the pulse, and the exponential
function can be expanded in the following series,
\begin{eqnarray}
&&a(z,t) = \sum_{m=-\infty}^{\infty} e^{im2k(z-p_0t/M)}  \times \notag \\
&& i^m \textit{J}_m \left(
\frac{\sqrt{2\pi}|d|^2E_1E_2}{\hbar^2\Delta}\tau
e^{-2\left(\frac{z}{c\tau}\right)^2}e^{-\left(\frac{p_0k\tau}{M}\right)^2/2}\right).
\label{eq5}
\end{eqnarray}
Note that this expression bears resemblance to the Raman-Nath solution obtained with constant laser
amplitudes \cite{theoryreview1,theoryreview2}. However, the specific Gaussian shape and an exponential
dependence on the duration parameter, $\tau$, of the Bessel-function argument highlights certain
specific physical properties of the diffraction process with Gaussian pulses. For all practical
purposes, the position argument, $z$, is much smaller than the pulse length $c\tau$ and the first
exponential term inside the Bessel-function will be replaced by unity in the following.  An example of
the results obtained from Eq.~(\ref{eq5}) for the momentum distribution, i.e., the absolute value of
the square of the Bessel-functions evaluated for different diffraction orders, $m$, are plotted in
Fig.~\ref{F1}(a) as function of the final state momentum $p=p_0+2m\hbar k$. Note the figure does not
show the time dependent diffraction, but the asymptotic momentum distribution after the pulse, for
different values of the pulse duration $\tau$.

The Bessel-functions with $m\neq 0$ all vanish for vanishing arguments, and we see that both short
($\tau\to 0$) and long ($\tau\to\infty$) pulses exclude diffraction to new momentum components. For
short pulses of given strength, the explanation is that there is simply no time to significantly
diffract the atoms, and the nearly vanishing $\tau$ argument suppresses the higher order Bessel
functions. For longer pulses the result can be ascribed to the exponential factor inside the
Bessel-functions, which in turn reflects the lack of energy conservation in the diffraction process:
In the rest frame of atoms moving at velocity $p_0/M$, the photons in the counter-propagating light
pulses are Doppler-shifted in frequency, causing an apparent change of $2\hbar k p_0/M$ in the field
energy per absorption-emission cycle involving both beams. This is a violation of energy conservation,
permitted because we consider a process of finite duration $\tau$. Note that the process is suppressed
by the exponential factor when the corresponding frequency mismatch exceeds $\propto 1/\tau$, in
support of our interpretation in terms of the energy-time Heisenberg uncertainty relation.
Alternatively, the finite duration pulses may be thought of as (Gaussian) frequency distributions of
light, and absorptions and emissions involving photons from the counter-propagating pulses may
conserve field energy in the frame moving at $p_0/M$, if the photons involved are at frequency
components shifted by the amount $\pm kp_0/M$ with respect to the field carrier frequency of the
pulses. The field power at those frequencies is just given by the Gaussian factor, appearing inside
the Bessel function arguments.

\subsection{Incorporation of the kinetic energy variation}

After having presented the equation governing the diffraction within
the Raman-Nath approximation, which does not take into
account the energy difference between states with different momenta,
it is natural to investigate if a simple modification of the final
result may capture the characteristic features of the energy
dispersion. To this purpose, we shall come back to our
interpretation of the last exponential factor inside the Bessel
function in Eq.~(\ref{eq5}). In the previous section we argued that this
quantity expresses the mean change of field energy $2\hbar k p_0/M$
per absorption emission cycle, experienced by the moving atom due to
the Dopper effect, and that it is appropriately measured relative to
the frequency witdth of the light pulses due to their finite duration.
For a Gaussian pulse, the frequency distribution is also Gaussian,
hence the Gaussian dependence inside the Bessel-function. Taking
into account that the different momentum states have different
kinetic energies, it is thus natural to suggest to apply the same
argument as above but taking into account the change of field energy
\emph{plus atomic kinetic energy}. In the frame moving at velocity
$p_0/M$ the atom is initially at rest. Scattering process into the $m$th
diffraction order by absorption and simulated emission of $m$
photons propagating in opposite directions changes the field energy
by an amount $2m\hbar kp_0/M$. Simultaneously atoms acquire a finite momentum kick of
$2m\hbar k$ in the moving frame, and hence their final kinetic energy is
$4m^2\hbar^2k^2/2M$. The total energy defect per absorption-emission event is therefore $2\hbar k(p_0+m\hbar k)/M$.
Since the pulses have Gaussian frequency widths
proportional to $1/\tau$, the atoms may interact with frequency
components shifted by $k(p_0+m\hbar k)/M$ in order to conserve
total (field plus atomic kinetic) energy.
Motivated by the above analysis we thus suggest to replace $p_0$ in the exponential inside the
Bessel-function in Eq.~(\ref{eq5}) by $p_0+m\hbar k$. This modification
leads to the expression,
\begin{eqnarray}
&&b(z,t) = \sum_{m=-\infty}^{\infty} e^{im2k(z-p_0t/M)}  \times \notag\\
&& i^m \textit{J}_m \left(
\frac{\sqrt{2\pi}|d|^2E_1E_2}{\hbar^2\Delta}\tau
e^{-\left(\frac{(p_0+m\hbar k)k\tau}{M}\right)^2/2}\right)
\label{eq6}
\end{eqnarray}
for different diffraction orders, $m$. The probability that an atom
with initial momentum $p_0$ acquires the final momentum $p_0+2m\hbar
k$, is $|J_m(f(p_0+m\hbar k))|^2,$
with
\begin{equation}
f(p)\equiv \frac{\sqrt{2\pi}|d|^2E_1E_2}{\hbar^2\Delta}\tau
e^{-\left(\frac{pk\tau}{M}\right)^2/2}.
\label{eq7}
\end{equation}

The results of this modified version of the Raman-Nath approximation
are shown in Fig.~\ref{F1}(b). We do not claim that such alteration is
an exact result or even a systematic expansion in any small
parameter, and although it is physically meaningful it must be
justified \emph{a posteriori}, for instance by comparison with an exact
calculation. Such assessment will be made in the following
section.

%\subsection{Time reversal symmetry and `detailed balance'}
%The expression (\ref{eq6}) for the state amplitudes suffers from an
%undesired asymmetry. The laser pulse is symmetric under time
%reversal, and hence the formal solution of the Schr\"odinger
%equation backwards in time from the end of the pulse to the start of the pulse
%should reproduce the same transition probability, $P$, between any pair of states as
%the equation for forward temporal propagation. This, in conjunction
%with unitarity of the time evolution operator, $U$, implies that the
%transition probability from momentum eigenstate $p_1$ to $p_2$ must equal
%the transition probability from $p_2$ to $p_1$,
%$|U_{p_1,p_2}|^2=|(U^\dagger)_{p_1,p_2}|^2=|(U_{p_2,p_1})^*|^2=|U_{p_2,p_1}|^2$.
%Our formula (\ref{eq6}) does not obey this symmetry, and we propose
%to restore the symmetry by the simplest possible modification,
%by taking the geometric mean of the `forward' and `backward'
%transfer probabilities, which leads to the expression
%\begin{equation}
%P(p_0\rightarrow p_0+2m\hbar k) = |J_m(f(p_0+2m\hbar
%k)J_{-m}(f(p_0))|,
%\label{eq8}
%\end{equation} with $f(p)$ defined in Eq.~(\ref{eq7}).
%The results of this expression are shown in Fig.~\ref{F1}(c).

\section{Results}

In the preceding section we have derived two different analytical solutions, which we shall
now compare to the exact numerical solution of the problem.

\subsection{Numerical solution}

The full Schr\"odinger equation without omission of the second
derivative term is amenable to numerical solution, and we have
carried out such computations for a wide range of atomic and field
parameters. Our numerical studies enable us to compare the
predictions of the analytical approximations to the exact numerical
solutions and assess their validity, and in particular, to
investigate whether the modified form Eq.~(\ref{eq6})
represent an improvement over the Eq.~(\ref{eq5}) or
not. To this end, we propagate wavefunctions, $d(z,t)$, which
interact with the Gaussian light pulses, in time from $t=-2\tau$
to $t=2\tau$ according to the Eq.~(\ref{eq2}). The initial
wave function represents a particle with a mean momentum, $p=p_0$,
and is modelled as a Gaussian wave packet
\begin{equation}
d(z,-\infty)=e^{-z^2 / 2\sigma^2} e^{ipz/\hbar}
\end{equation}
where the waist, $\sigma=10/k$, is chosen wide enough that the
momentum spread of the initial wave packet is narrow in comparison
to the resulting momentum separation of adjacent diffraction orders.

For further convenience, following Ref.~\cite{Keller1999a}, we
introduce dimensionless parameters for time and laser field
intensity
\begin{equation}
q = \left|\frac{d^2E_1E_2}{4\hbar^2\omega_r\Delta}\right| ; \;\;\;\;\;\;\;\;\;\; \tau_r=\omega_r t,
\end{equation}
where $\omega_r=\hbar k^2/2M$ is the photon recoil frequency.

\begin{figure}[t!] %  figure placement: here, top, bottom, or page
\centering
\includegraphics[width=8.6cm]{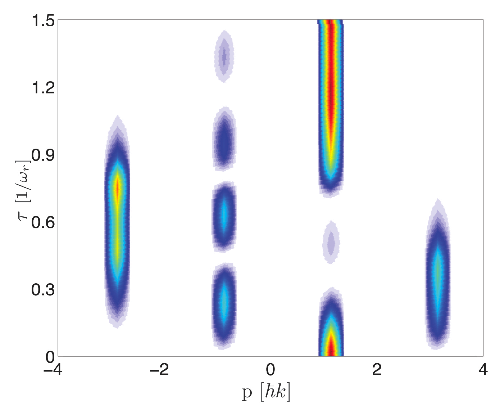}
\caption{(Color online) Diffraction patterns for various pulse durations, $\tau$, as predicted by the modified Raman-Nath solution Eq.~(\ref{eq6}), plotted as a function of the momentum, $p$. The color indicates the amplitude such that red corresponds to the maximum and dark blue to the minimum. The pulse intensity $q=3.5$ and the atom arrives to the interaction region with an initial momentum $p=\hbar k$.}
\label{F3}
\end{figure}
\begin{figure}[t!] %  figure placement: here, top, bottom, or page
\centering
\includegraphics[width=8.6cm]{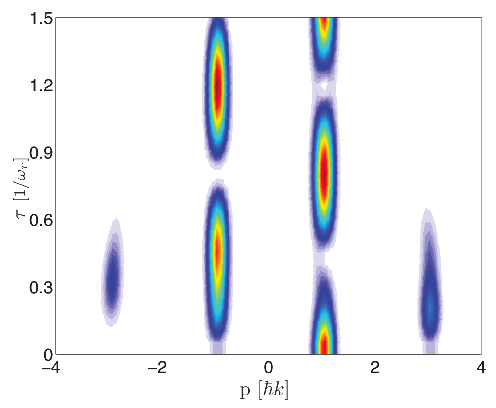}
\caption{(Color online) Diffraction patterns for various pulse durations, $\tau$, as predicted by the full numerical solution to Eq.~(\ref{eq2}), plotted as a function of the momentum, $p$. The color indicates the amplitude such that red corresponds to the maximum and dark blue to the minimum. The pulse intensity $q=3.5$ and the atom arrives to the interaction region with an initial momentum $p=\hbar k$.}
\label{F4}
\end{figure}

\begin{figure}[t!] %  figure placement: here, top, bottom, or page
\centering
\includegraphics[width=8.6cm]{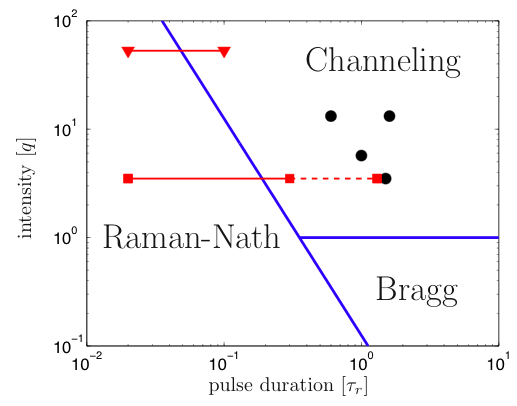}
\caption{(Color online) Pulse duration versus intensity diagram for atomic diffraction dividing the parameter space
in three distinctively different regimes. Black bullets correspond to the experiments in Ref.~\cite{Keller1999a}.
Triangles joined by a line correspond to the modified Raman-Nath solution in Fig.~\ref{F1}(b) and the boxes joined
by a line are from Fig.~\ref{F3}. Dashed line marks the region where the modified Raman-Nath formula loses its validity.}
\label{F5}
\end{figure}

\subsection{Comparison of different analytical solutions}

In Fig.\ref{F1}(a)-(c), we have plotted the diffraction patterns for a
range of pulse durations, $\tau$, as functions of the momentum, $p$
for two different approximations and the exact solution to the
diffraction problem, respectively. We obtain the diffraction peak intensities
as a square modulus of the Fourier transformed wave functions.
Note that the `onion shape' does not show the
momentum distribution as a function of time, instead, each horizontal cut
through the figure indicates the asymptotic momentum distribution
after a pulse of the given duration. The pulse parameters for all of
the frames are $q=52.9$ and $p_0=20\hbar k$. The modified Raman-Nath solution, shown in Fig.~\ref{F1}(b),
is observed to be in an excellent agreement
with the exact numerical solution, Fig.~\ref{F1}(c), for these parameters .

To obtain better quantitative feeling for the quality of the
equation (\ref{eq6}), we plot the diffraction pattern for a single pulse
duration, $\tau=0.0375 /\omega_r$ as shown in Fig.~\ref{F2}. Here the red
diamonds, blue boxes and black bullets correspond to the values
shown in the frames (a), (b), and (c) in Fig.~\ref{F1},
respectively. From this graph, it is clear that our modified
solution still works excellently in the regime where the Raman-Nath
approximation fails.

\subsection{Validity assessment and relation to experiments}

We have also compared our modified Raman-Nath prediction, Eq.~(\ref{eq6}),  with the actual experiment
as reported in Ref.~\cite{Keller1999a} in addition to the full numerical solution. In the experiment,
the smooth-shaped pulse was varied as $\cos^2(\alpha t)$ which is well approximated by our Gaussian
shaped pulse and allows a direct comparison. With parameters corresponding to the experiment, $q=3.5,
\tau_r=[0,1.5]$ and $p_0=\hbar k$, Fig.~\ref{F3} shows our prediction while Fig.~\ref{F4} is the
corresponding numerical solution. Comparison between these and the experimental result presented in
\cite{Keller1999a}, we observe all of them to be in fair agreement for short pulse durations,
\emph{i.e.}, in the Raman-Nath regime. For longer pulse durations, both experiment and numerical
solution exhibits Bragg-like channeling behavior with an associated Rabi cycling. Such long-time
behavior for these parameters cannot be reproduced by our analytical formulas. This may be because the
experiment is carried out with parameters very far from the Raman-Nath regime. The experiment is in
fact not that far from the Bragg regime, where we know that a correct treatment of the kinetic energy
is crucial.

Figure~\ref{F5} is in parts reproduced from Ref.\cite{Keller1999a}
and summarizes the applicability of various approximations on the
$(\tau_r,q)$ parameter space. In the spirit of Ref.~\cite{Keller1999a},
the region boundaries are defined by $\tau_r=1/2\sqrt{2q}$ and $q=1$.
Black dots mark the experiments as in Ref.~\cite{Keller1999a}. The upper horizontal line connecting triangle
symbols corresponds to the results shown in Fig.~\ref{F1}(b) and the
lower line connecting square symbols depicts the parameters used in
Fig.~\ref{F3}. The dashed part of the lower line indicates that Eq.~(\ref{eq6})
fails for these parameters, deep into the
channeling regime. However, it is clear that the validity of our
analytical solution enters the previously unknown channeling territory.
It is also worth noting that for the case of high intensity
(high $q$ value), demonstrated in Fig.~\ref{F1}, our expression maintains its validity for all pulse
durations, $\tau$, since the diffraction process itself is confined
to relatively short times.

\section{Discussion}
We have presented an analytical solution to a problem of atomic
diffraction from a pulsed time-dependent light grating for the so
called Raman-Nath regime where the kinetic energy of the atoms
participating in the diffraction process may be neglected. We have
extended this approximation by including certain aspects of the
kinetic energy in the diffraction process and we have proposed a
phenomenological improvement to this solution which is based on
physically motivated arguments. Through a comparison
with an exact numerical solution of the Schr\"odinger equation, it
is shown that our modification extends the validity regime of the
parameter space of the analytical solution from the Raman-Nath into
the channeling regime. Qualitatively the modification predicts the
asymmetry due to the Doppler effect in the diffraction pattern. The
proposed analytical formula is compared with available numerical results
and is found to be in good agreement even fairly deep in the
channeling regime of strong and long interactions. Although,
containing certain elements of Bragg diffraction, the proposed equation is not,
however, capable of predicting correctly the effective two-state
Rabi cycling between momentum states $\pm p_0$. Exploration of the
channeling regime from the Bragg-formalism seems a promising avenue
for attempts to provide good analytical expressions for atomic
diffraction closer to the Bragg regime. Together with the present
work, this may perhaps exhaust the full parameter range with good
analytical approximations for the atomic diffraction.

The authors acknowledge support from a NATO linkage grant, and A. M. acknowledges support via Yerevan
State University grant NFSAT/CRDF UCEP-02/07.


\begin{thebibliography}{90}
\bibitem{earlyref1} Phillip L. Gould, George A. Ruff, and David E. Pritchard,
Phys. Rev. Lett. \textbf{56}, 827 - 830 (1986).
\bibitem{earlyref2} O. Carnal and J. Mlynek, Phys. Rev. Lett. 66, 2689
(1991).
\bibitem{earlyref3} D.W. Keith, C. R. Ekstrom, Q. A. Turchette, and D. E. Pritchard, Phys. Rev. Lett.
\textbf{66}, 2693 (1991).
\bibitem{earlyref4} F. Riehle, Th. Kister, A. Witte, J. Helmcke, and Ch. Bord\'e, Phys. Rev.
Lett. \textbf{67}, 177 (1991).
\bibitem{earlyref5} F. Shimizu, K. Shimizu, and H. Takuma, Phys. Rev. A \textbf{46}, R17
(1992).
\bibitem{earlyref6} M. Kasevich and S. Chu, Phys. Rev. Lett. \textbf{67}, 181 (1991).
\bibitem{useofdiff1} J. M. McGuirk, G. T. Foster, J.B. Fixler, M. J. Snadden, and M. A. Kasevich, Phys. Rev.
A \textbf{65}, 033608 (2002).
\bibitem{useofdiff2} T. L. Gustavson, P. Bouyer, and M. A. Kasevich, Phys. Rev. Lett.
\textbf{78}, 2046 (1997).
\bibitem{useofdiff3} A. Lenef, T. D. Hammond, E. T. Smith, M. S. Chapman, R. A. Rubenstein, and
D. E. Pritchard, Phys. Rev. Lett. \textbf{78}, 760 (1997).
\bibitem{useofdiff4} B. Canuel, F. Leduc, D. Holleville, A.
Gauguet, J. Fils, A. Virdis, A. Clairon, N. Dimarcq, Ch. J. Bord\'e, A. Landragin, and P. Bouyer,
Phys. Rev. Lett. \textbf{97}, 010402 (2006).
\bibitem{manybody1} J. Stenger, S. Inouye, A. P. Chikkatur, D. M. Stamper-Kurn, D. E. Pritchard, and W. Ketterle,
Phys. Rev. Lett. \textbf{82}, 4569 (1999).
\bibitem{manybody2} Yu. B. Ovchinnikov, J. H. M\"uller, M. R. Doery, E. J. D.
Vredenbregt, K. Helmerson, S. L. Rolston, and W. D. Phillips, Phys. Rev. Lett. {\bf 83}, 284 (1999).
\bibitem{manybody3} J. Steinhauer, R. Ozeri, N. Katz, and N. Davidson, Phys. Rev. Lett. 88, 120407 (2002).
\bibitem{deposition} B. Rohwedder, Am. J. Phys. \textbf{75}, 394 (2007).
\bibitem{Andersen2007a}
M.~F. Andersen, C. Ryu, P. Clad\'e, V. Natarajan, A. Vaziri, K.
Helmerson and W.~D. Phillips, \emph{Phys. Rev. Lett.} {\bf 97},
170406 (2006).
\bibitem{Simula2007a}
T.~P. Simula, N. Nygaard, S.~X. Hu, L.~A. Collins, B.~I. Schneider,
and K. M\o lmer, \emph{cond-mat/1234.5678} (2007).
\bibitem{theoryreview1} C. S. Adams, M. Siegel, and J. Mlynek.
Physics Reports, \textbf{240}, 143 (1994).
\bibitem{theoryreview2} Atom Interferometry, edited by R.
Paul Berman (Academic, London, 1997).
\bibitem{Keller1999a}
C. Keller, J. Schmiedmayer, A. Zeilinger, T. Nonn, S. D\"urr, and G.
Rempe, \emph{Appl.Phys. B} {\bf 69}, 303 (1999).
\bibitem{Muller2007a}
H. M\"uller, S. Chiow, and S. Chu, arXiv:0704.2627 (2007).
\bibitem{Muradyan}
V.M.Arutunyan and A.Zh.Muradyan, Dokladi Akad. Nauk Arm. SSR, \textbf{60}, n.5, 275 (1975).

\end{thebibliography}
\end{document}